\documentclass[a4paper, 12pt]{article}
\usepackage{graphicx}
\usepackage[small]{subfigure,epsfig}
\usepackage {amsmath}
\usepackage{amssymb}
\usepackage{cite}
\usepackage{indentfirst}
\usepackage{array,longtable,lscape}
\numberwithin{equation}{section} \numberwithin{table}{section}
\usepackage{lscape}

\begin{document}

\title{Self-organization of adiabatic shear bands in OFHC copper and HY-100 steel}
\author{N.A. Kudryashov, P.N. Ryabov, A.S. Zakharchenko}
\date{\small{Department of Applied Mathematics, National Research Nuclear University MEPHI, 31 Kashirskoe Shosse, 115409 Moscow, Russian Federation}}

\maketitle

\begin{abstract}
In this paper we study the self-organization process of adiabatic shear bands in OFHC copper and HY-100 steel taking into account strain hardening factor. Starting from mathematical model we present new numerical approach, which is based on Courant -- Isaacson -- Rees scheme that allows one to simulate fully localized plastic flow. To prove accuracy and efficiency of our method we give solutions of two benchmark problems. Next we apply the proposed method to investigate such quantitative characteristics of self-organization process of ASB as average stress, temperature, localization time and distance between ASB. Then we compare the obtained results with theoretical predictions by other authors.
\end{abstract}

\emph{Keywords:} Adiabatic shear banding; Self-organization; Numerical simulation; Courant -- Isaacson -- Rees scheme.

PACS 83.50.-v, 91.55.Mb

\section{Introduction}
It is known that one of the most curious phenomenon in nonlinear science is a self-organization of stable structures in physical systems. A striking example of such processes is a process of self-organization of adiabatic shear bands in ductile materials. In the last few decades this phenomenon attracts increased attention since it turned out that ASB is one of the major failure mechanisms in variety of materials exposed to high strain-rates deformations. The localization places of ASB are more brittle than the surrounding areas so that in the case of prolonged deformation the material is destroyed along the ASB. In fact, adiabatic shear bands are narrow regions (1-500 $\mu$m) where the high temperatures and deformations are reached due to the conversion of plastic work to heat without heat transfer. The following phenomenon was observed in a series of technological processes of manufacturing, military and space industry. Examples of such processes are found in shock loading, metal forming, ballistic impact \cite{Schneider_2004,Seidel2001, Moss1981} and has even been the hypothesis of the space shuttle crash \cite{Shockey2007, Rittel2005}.

For the first time the self-organization process of adiabatic shear bands was observed in experimental works of Nesterenko et.al. \cite{Nesterenko1995, Nesterenko1998}. Authors studied the radial collapse of thick-walled cylinder under controlled explosion technique. They succeeded to reach a high initial strain-rate deformation $\sim 10^4$ s$^{-1}$. Nesterenko et.al. observed the formation of multiple shear bands spaced periodically with a characteristic distance between them. To compare those results with theoretical predictions they used Grady and Kipp \cite{Grady-Kipp} and Wright--Oscendon estimates \cite{Wright-Ockendon} of the distance between ASB. According to Grady and Kipp the average distance between ASB ($L_{GK}$) is governed by momentum diffusion during the growth time \cite{Grady-Kipp}. To calculate the distance between ASB Wright-Ockendon have used perturbation approach \cite{Wright-Ockendon}. Authors showed that for different Fourier modes the growth rate varies but there is finite wavelength with the maximum growth rate which corresponds to the most probable spacing between shear bands ($L_{WO}$) \cite{Wright-Ockendon}. It should be noted that the same method was used in works by Molinari \cite{Molinari}, Batra and his coauthors \cite{BatraChen1999, BatraChen2000, BatraWei2006, BatraWei2007} to estimate the spacing between ASB for the different materials,  models and constitutive relationships. Among other works devoted to investigation of the features of adiabatic shear bands collective behavior we have to note the works by Zhou et.al. \cite{Zhou2006a, Zhou2006b}. Based on the method of characteristic, authors proposed the numerical algorithm for studying the ASB formation in one-dimension. Using the combined dimensional analysis and numerical simulation they obtained the empirical formula for the estimation of band spacing in materials ($L_{ZWR}$). However all of these theoretical estimates $L_{WO}, L_{GK}, L_{ZWR}$ do not take into account the strain hardening effect. In work \cite{BatraWei2006} authors have derived the formula for describing the strain-hardening materials but this estimation is valid on initial stage of multiple ASB formation. Thus, motivated by \cite{Zhou2006b} and \cite{BatraWei2006}, we are going to study the self--organization process of adiabatic shear bands taking into account the strain hardening effect in OFHC copper and HY-100 steel.

Besides the experimental approach the most effective way to study the process of ASB formation is a numerical simulation. There are a lot of works where the method of numerical simulation was successfully applied \cite{DiLellio1997, DiLellio1998,DiLellio2003, Bayliss1994, Doridon2004, Edwards1998}. However most of them concentrated on the investigation of a single band evolution. Many of these numerical algorithms designed to solve a particular problem and it is difficult to generalize them to solve the problems of another type. In work \cite{Zhou2006b} authors noted that in the case of multiple shear bands formation at the random places these algorithms lose effectiveness. To overcome this difficulty we developed new numerical methodology that allows one to solve the one--dimensional model of adiabatic shear bands formation.

Our work is organized as follows. In the section 2  we present the basic equations for the description of the ASB formation in one--dimension. In section 3 we describe our numerical methodology. In section 4 we solve two benchmark problems from\cite{Walter1992} and finally in section 5 we present the results of our numerical investigation of the multiple ASB formation taking into account strain hardening factor.

\section{One-dimensional model of adiabatic shear bands formation}
\begin{figure}[!htb]
\begin{center}
\includegraphics[width=13 cm]{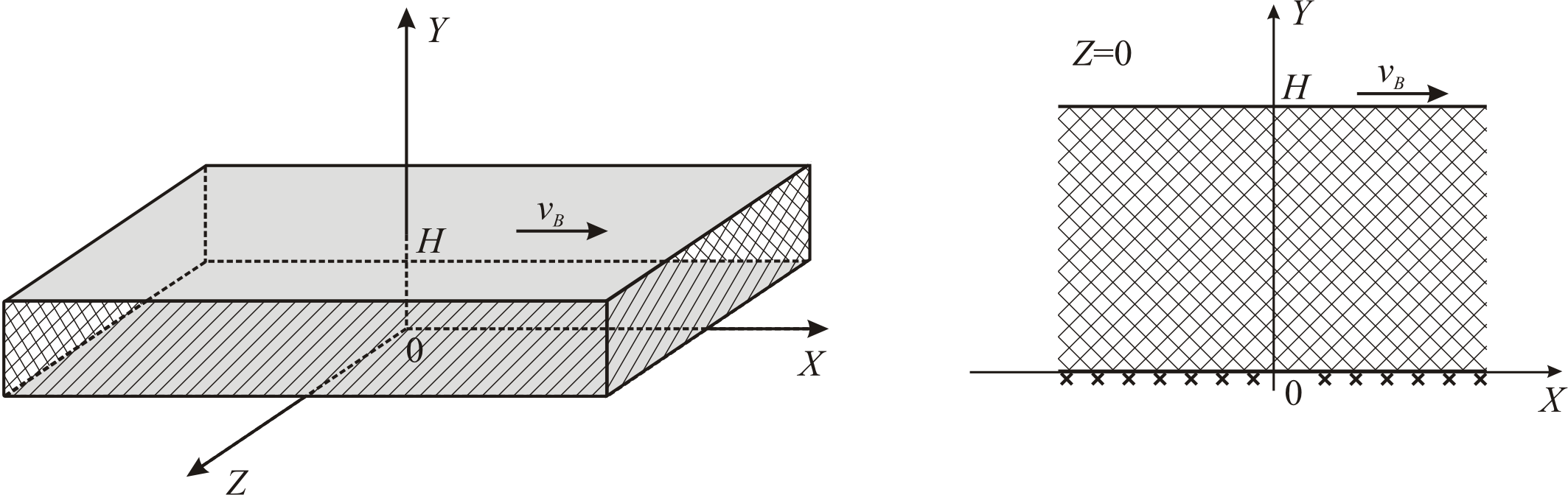}
\end{center}
\caption{Geometry of the problem}\label{fig:1}
\end{figure}

We consider the process of shear deformation in an infinite slab of incompressible elastic, thermo-visco-plastic material. Slab height is $0\leq y\leq H$. The geometry of the problem is presented in Fig. \ref{fig:1}.

According to \cite{Walter1992, Wright1985, Wright1987} the problem considered can be expressed by the following system of equations
\begin{equation}\label{Eq1}
v_t=\frac{1}{\rho}\tau_y,
\end{equation}
\begin{equation}\label{Eq2}
\tau_t-\mu v_y=-\mu \dot{\varepsilon}^p,
\end{equation}
\begin{equation}\label{Eq3}
\psi_t=\frac{\tau \dot{\varepsilon}^p}{\kappa(\psi)} ,
\end{equation}
\begin{equation}\label{Eq5}
\dot{\varepsilon}^p=\Phi^{-1}(T,\psi,\tau).
\end{equation}
\begin{equation}\label{Eq6}
C \rho T_t= \left(k T_y\right)_y+\beta \tau \dot{\varepsilon}^p,
\end{equation}
where $y$ is the Lagrangian spatial coordinate, $t$ is the time, $\tau(y,t)$ is the nonzero component of the stress tensor, $v(y,t)$ is the velocity, $T(y,t)$ presents the temperature of material, $\dot{\varepsilon}^p(y,t)$ is the plastic strain rate. Furthermore, $\rho$ is a mass density, $\beta$ is the Taylor-Quinney parameter, $\mu(T), C(T)$ and $k(T)$ -- elastic shear modulus, specific heat and thermal conductivity respectively that can be a functions of temperature. Here, Eq. \eqref{Eq1} is the equation of motion, Eq. \eqref{Eq2} is the elastic relation, Eq. \eqref{Eq3} describes the strain hardening process, Eq. \eqref{Eq5} is the plastic flow law and Eq. \eqref{Eq6} is the energy conservation. Note that in works \cite{Zhou2006a,Zhou2006b, JohnsonCook} to define the value of plastic strain $\varepsilon^p(y,t)$ authors used the relation
\begin{equation} \label{Eq4}
\varepsilon^p=\int_0^t \dot{\varepsilon}^p dt',
\end{equation}
that is the equivalent to the strain hardening variable $\psi$ \cite{Walter1992}. We use both relations \eqref{Eq3} and \eqref{Eq4} to calculate the plastic strain and then compare obtained results for $\psi$ and $\varepsilon^p$ with each other.

It is known that the process of initiation and formation of  adiabatic shear bands accompanied by large deformations and temperature changes. Thus the phase changes may take place. However, we ignore such effects.

Initially, the slab is free of strain and is shearing under the uniform strain rate $\dot{\varepsilon}_0$. The initial conditions, in that case, take the form
\begin{equation} \label{Eq7}
\begin{gathered}
v(y,0)=v_0(y), \quad T(y,0)=T_0(y), \quad \tau(y,0)=\tau_0(y),\\
\psi(y,0)=\psi_0, \quad \dot{\varepsilon}^p(y,0)=\dot{\varepsilon}_0, \quad \varepsilon(y,0)=0.
\end{gathered}
\end{equation}

Boundary conditions were used in the form
\begin{equation} \label{Eq8}
\begin{gathered}
v(0,t)=0, \quad v(H,t)=v_B, \quad
\frac{\partial T}{\partial y}(y=0)=0, \quad \frac{\partial T}{\partial y}(y=H)=0.
\end{gathered}
\end{equation}
Note that one can use other boundary conditions, example of which is given in \cite{Zhou2006a, Zhou2006b, Batra2006, KKR2011, Koshkin2010}.

\section{Numerical strategy}

Let us consider the process of adiabatic shear band formation described by the closed system of Eqs. \eqref{Eq1}--\eqref{Eq6} taking into account initial and boundary conditions \eqref{Eq7}, \eqref{Eq8}. It is known that effective methods to solve such initial--boundary value problem are numerical methods. In order to construct a numerical solution, we introduce a rectangular grid $\Pi=\{y_{j}=j\Delta y, t^{n}=n\Delta t\}$, where $j=0...J, n=0...N$, $\Delta y, \Delta t$ are integration space and time steps respectively. So we introduce the following notations of functions in the grid nodes $f(y_j,t^n)=f^n_j$.

The system of Eqs. \eqref{Eq1}--\eqref{Eq6} is mixed. It contains the hyperbolic subsystem of Eqs. \eqref{Eq1} -- \eqref{Eq2}, parabolic Eq. \eqref{Eq6}, nonlinear ordinary differential Eq. \eqref{Eq3} and nonlinear relation \eqref{Eq5}. To obtain the numerical solution of that system we have to take into account its mixed type.

At first we solve the mechanical part of the problem and determine stress $\tau$ and velocity $v$. For this purpose we consider the subsystem of hyperbolic equations
\begin{equation} \label{eq:1.31}
\begin{gathered}
v_t - \frac{1}{\rho}\tau_y=0,\\
\tau_t-\mu v_y=-\mu \dot{\varepsilon}^p.
\end{gathered}
\end{equation}
Note that shear modulus $\mu$ can be the function of temperature $T$.

To construct a finite-difference scheme for approximating \eqref{eq:1.31}, we have used the scheme of Courant-Isaacson-Rice type (CIR) \cite{Chm_Kulikovskii}. Suppose we have a system of hyperbolic equations written in matrix form
\begin{equation} \label{eq:1.32}
\mathbf{U}_t+A(\mathbf{U})\mathbf{U}_y=0, \quad A(\mathbf{U})=\Omega_R\Lambda\Omega_L, \quad \mathbf{U}=(u_1,u_2)^T,
\end{equation}
where $\mathbf{U}$ is the vector that depends on variables $y$, $t$. The diagonal matrix $\Lambda$ consists of eigenvalues of the matrix $A$, i.e. $\Lambda=\text{diag}[\lambda_1, \lambda_2]$. The matrix $\Omega_R$ and $\Omega_L$ are connected by relation $\Omega_R\Omega_L=\text{I}$, where $\text{I}$ is the identity matrix.
In this case the finite-difference scheme of CIR's type is represented as
\begin{equation} \label{eq:1.33}
\begin{gathered}
\frac{\mathbf{U}^{n+1}_{j}-\mathbf{U}^{n}_{j}}{\Delta t}+A(\mathbf{U}^n_j)\frac{\mathbf{U}_{j+1/2}-\mathbf{U}_{j-1/2}}{\Delta y}=0,\\
\mathbf{U}_{m+1/2}=\frac{1}{2}(\mathbf{U}^{n}_{m}+\mathbf{U}^{n}_{m+1})+\frac{1}{2}S^n_j(\mathbf{U}^{n}_{m}-\mathbf{U}^{n}_{m+1}), \quad m=j, j-1\\
S=\Omega_R[\text{sign}\lambda_p\delta_{pl}]\Omega_L.
\end{gathered}
\end{equation}

We apply the finite-difference scheme \eqref{eq:1.33} to the system \eqref{eq:1.31} without its right part. For this we take $\mathbf{U}=(v,\tau)^{T}$ and get
\begin{equation} \label{eq:1.35}
\begin{gathered}
A=\begin{pmatrix}
0 & -\frac{1}{\rho} \\
-\mu & 0
\end{pmatrix}
, \quad
\Lambda=\begin{pmatrix}
\sqrt{\frac{\mu}{\rho}} & 0\\
0 & -\sqrt{\frac{\mu}{\rho}}
\end{pmatrix},\\
\Omega_R=\begin{pmatrix}
1 & 1\\
-\sqrt{\mu \rho} & \sqrt{\mu \rho}
\end{pmatrix}
, \quad
\Omega_L=\begin{pmatrix}
\frac{1}{2} & -\frac{1}{2\sqrt{{\mu}{\rho}}}\\
\frac{1}{2} & \frac{1}{2\sqrt{{\mu}{\rho}}}
\end{pmatrix}.
\end{gathered}
\end{equation}
If we convert the matrix form of finite-difference scheme to the notation for each equation separately and take into account its right part we will obtain the following system of equations
\begin{equation} \label{eq:1.36}
\begin{gathered}
v^{n+1}_{j}=v^{n}_{j}+\frac{\Delta t}{2\rho \Delta y}(\tau^{n}_{j+1}-\tau^{n}_{j-1})+\frac{1}{2}\sqrt{\frac{\mu^n_j}{\rho}}\frac{\Delta t}{\Delta y}(v^{n}_{j+1}-2v^{n}_{j}+v^{n}_{j-1}),
\end{gathered}
\end{equation}
\begin{equation} \label{eq:1.37}
\begin{gathered}
\tau^{n+1}_{j}=\tau^{n}_{j}+\frac{\mu^n_j \Delta t}{2\Delta y}(v^{n}_{j+1}-v^{n}_{j-1})+\frac{1}{2}\sqrt{\frac{\mu^n_j}{\rho}}\frac{\Delta t}{\Delta y}(\tau^{n}_{j+1}-2\tau^{n}_{j}+\tau^{n}_{j-1})-\\
-\mu^n_j\Delta t\frac{\dot{\varepsilon}_{j}^{p(n+1)}+\dot{\varepsilon}_{j}^{p(n)}}{2},
\end{gathered}
\end{equation}
where $j=1,...,J-1$. State Eq. \eqref{Eq5} closes the system \eqref{eq:1.36}-\eqref{eq:1.37}. We write Eq. \eqref{Eq5} in a discrete form
\begin{equation} \label{eq:1.38}
\dot{\varepsilon}_{j}^{p(n+1)}=\Phi^{-1}(T^{n}_{j},\psi^{n}_{j},\tau^{n+1}_{j}).
\end{equation}

Substituting Eq.\eqref{eq:1.38} in Eq.\eqref{eq:1.37}, we obtain the nonlinear algebraic equation for the unknown quantity $\tau^{n+1}_{j}$
\begin{equation} \label{eq:1.39}
\tau^{n+1}_{j}=\tilde{\tau}^{n}_{j}-\mu^n_j\Delta t\frac{\Phi^{-1}(T^{n}_{j},\psi^{n}_{j},\tau^{n+1}_{j})+\Phi^{-1}(T^{n}_{j},\psi^{n}_{j},\tau^{n}_{j})}{2},
\end{equation}
where
\begin{equation} \label{eq:1.40}
\tilde{\tau}^{n}_{j}=\tau^{n}_{j}+\frac{\mu^n_j\Delta t}{2\Delta y}(v^{n}_{j+1}-v^{n}_{j-1})+\frac{1}{2}\sqrt{\frac{\mu^n_j}{\rho}}\frac{\Delta t}{\Delta y}(\tau^{n}_{j+1}-2\tau^{n}_{j}+\tau^{n}_{j-1}).
\end{equation}
To solve Eq.\eqref{eq:1.39} with respect to $\tau^{n+1}_{j}$, we have used the Newtons method, because the simple iteration method does not converge at the moment of the adiabatic shear band formation.

The finite-difference scheme \eqref{eq:1.36}-\eqref{eq:1.37} is sustainable under the condition
\begin{equation} \label{eq:1.45}
\max_{j,n}\sqrt{\left (\frac{\mu^n_j}{\rho}\right )}\frac{\Delta t}{\Delta y}\leqslant 1.
\end{equation}
In this regard, the calculations will be carried out using the condition
\begin{equation} \label{eq:1.53}
\Delta t=\min_{0\leq j \leq J}\alpha\Delta y \sqrt{\frac{\rho}{\mu^n_j}},
\end{equation}
where $\alpha$ is any value at the interval $(0,1]$.

A magnitude of plastic strain ${\varepsilon}_{j}^{p(n+1)}$ is calculated using the trapezoidal rule as
\begin{equation} \label{eq:1.57}
{\varepsilon}_{j}^{p(n+1)}={\varepsilon}_{j}^{p(n)}+\frac{\Delta t}{2}\left( \dot{\varepsilon}_{j}^{p(n+1)}+\dot{\varepsilon}_{j}^{p(n)}\right).
\end{equation}

To construct the numerical approximation of the strain hardening Eq. \eqref{Eq3} with the initial condition \eqref{Eq7}, we use the fourth order Runge-Kutta method. According to this method we obtain the discrete equation for variable $\psi$
\begin{equation} \label{eq:1.46}
\begin{gathered}
\psi^{n+1}_{j}=\psi^{n}_{j}+\frac{\Delta t}{6}(k_1+2k_2+2k_3+k_4),\\
k_1=f(\tau^{n}_{j},\dot{\varepsilon}^{p(n)}_{j},\psi^{n}_{j}), \\
k_2=f\left(\tau^{n+1/2}_{j}, \dot{\varepsilon}^{p}\left(T^{n}_{j},\tau^{n+1/2}_{j},\psi^{n}_{j}+\frac{\Delta t}{2}k_1 \right), \psi^{n}_{j}+\frac{\Delta t}{2}k_1\right),\\ k_3=f\left(\tau^{n+1/2}_{j}, \dot{\varepsilon}^{p}\left(T^{n}_{j},\tau^{n+1/2}_{j},\psi^{n}_{j}+\frac{\Delta t}{2}k_2 \right), \psi^{n}_{j}+\frac{\Delta t}{2}k_2\right),\\
k_4=f\left(\tau^{n+1}_{j}, \dot{\varepsilon}^{p}\left(T^{n}_{j},\tau^{n+1}_{j},\psi^{n}_{j}+{\Delta t} k_3 \right), \psi^{n}_{j}+{\Delta t} k_3\right),
\end{gathered}
\end{equation}
where $\tau_j^{n+1/2}=(\tau_j^{n+1}+\tau_j^{n})/2$ and $f$ has the form
\begin{equation} \label{eq:1.47}
f(\tau,\dot{\varepsilon}^{p},\psi)=\frac{\tau\cdot\dot{\varepsilon}^{p}}{\kappa(\psi)}.
\end{equation}
For the approximation of heat equation \eqref{Eq6}, we use a two-layer implicit finite-difference scheme. As a result we obtain
\begin{equation} \label{eq:1.49}
\begin{gathered}
\frac{T^{n+1}_{j}-T^{n}_{j}}{\Delta t}=\frac{{k}^{n+1}_{j+1}+k^{n+1}_{j}}{2C^{n+1}_{j}\rho}\frac{T^{n+1}_{j+1}-T^{n+1}_{j}}{{\Delta y}^{2}}-\frac{{k}^{n+1}_{j}+k^{n+1}_{j-1}}{2C^{n+1}_{j}\rho}\frac{T^{n+1}_{j}-T^{n+1}_{j-1}}{{\Delta y}^{2}}+\\
+\frac{\beta}{C^{n+1}_{j}\rho}\frac{(\tau^{n+1}_{j}+\tau^{n}_{j})(\dot{\varepsilon}_{j}^{p(n+1)}+\dot{\varepsilon}_{j}^{p(n)})}{4}.
\end{gathered}
\end{equation}
The system of algebraic equations \eqref{eq:1.49} is solved using the sweep method.

The above scheme is applied to all internal nodes ($1\leqslant j\leqslant J-1$). The two boundary nodes ($j=0$ or $j=J$) are processed separately. For constant velocity and adiabatic boundary conditions \eqref{Eq8}, the boundary temperatures are calculated after the internal nodes are solved
\begin{equation} \label{eq:1.52}
\begin{gathered}
\tau^{n+1}_{0}=\tau^{n}_{0}+{\Delta t}\mu^n_j\left(\frac{v^{n+1}_{1}-v^{n+1}_{0}}{\Delta y}\right)-{\Delta t}\mu^n_j\left(\frac{\dot{\varepsilon}_{0}^{p(n+1)}+\dot{\varepsilon}_{1}^{p(n)}}{2}\right),\\
\tau^{n+1}_{J}=\tau^{n}_{J}+{\Delta t}\mu^n_j\left(\frac{v^{n+1}_{J}-v^{n+1}_{J-1}}{\Delta y}\right)-{\Delta t}\mu^n_j\left(\frac{\dot{\varepsilon}_{J}^{p(n+1)}+\dot{\varepsilon}_{J-1}^{p(n)}}{2}\right),\\
v^{n+1}_{0}=0, \quad
v^{n+1}_{J}=\dot{\varepsilon}_0H,\\
T^{n+1}_{0}=T^{n+1}_{1}, \quad T^{n+1}_{J}=T^{n+1}_{J-1}.
\end{gathered}
\end{equation}
Because the boundary conditions are given implicitly by the terms $\dot{\varepsilon}_{0}^{p(n+1)}$ and $\dot{\varepsilon}_{J}^{p(n+1)}$ we use iterations for them.

\section{Verification of the numerical methodology}

Let us present the verification results of our numerical methodology. We consider two benchmark problems which were solved by Walter in \cite{Walter1992}. It should be noted that some of these problems were considered in works \cite{DiLellio1997, DiLellio1998, Zhou2006a}.

\subsection{Benchmark problem 1: Adiabatic shear band formation in steel HY-100 without strain hardening effects}

Let us discuss the first benchmark problem that was studied in \cite{Walter1992, Zhou2006a}.

The slab is 6.94 mm long, deforming at nominal strain rate of $\dot{\varepsilon}_0 = 750$ s$^{-1}$. The plastic flow law was used in the form
\begin{equation} \label{eq:1.20}
\tau=\kappa_0 g(T){\left(1+\frac{\dot{\varepsilon}^p}{\dot{\varepsilon}_y}\right)}^{m},
\end{equation}
where $\kappa_0$ is the static yield stress, $\dot{\varepsilon}_y$ is a reference strain rate, $m$ is the strain-rate sensitivity and $g(T)$ is a thermal softening factor that was used in the form
\begin{equation} \label{eq:1.20_1}
g(T)=\exp(-aT).
\end{equation}
The process of strain-hardening was neglected in \eqref{eq:1.20}.

According to the equation \eqref{eq:1.20} we obtain the expression for $\dot{\varepsilon}^p$ as
\begin{equation} \label{eq:1.21}
\dot{\varepsilon}^p=\frac{\dot{\varepsilon}_y}{2}\left( {\left[ \frac{|\tau|}{\kappa_0g(T)}\right]}^{\frac{1}{m}}-1 \right)\left( 1+\mbox{sign}\left[ {\left[ \frac{|\tau|}{\kappa_0g(T)}\right]}^{\frac{1}{m}}-1 \right] \right).
\end{equation}
Material parameters in this case are presented in Table \ref{tabl1} and are identical to those used by Walter \cite{Walter1992}.
\begin{table}[h]
  \footnotesize
  \caption{Elastic, thermo-visco-plastic parameters of HY-100 steel.}\label{tabl1}
  \begin{center}
    \begin{tabular}{c|c|c|c|c|c|c|c|c|c}
        \hline
$\mu\mbox{,}$ & $\rho\mbox{,}$ & $k\mbox{,}$ & $C\mbox{,}$ &${\dot{\varepsilon}_y}\mbox{,} $&$a\mbox{,}$&$m$&$\kappa_0$,& $\psi_0$&$n$\\
$\mbox{GPa}$&$\mbox{kg}/\mbox{m}^3$&$\mbox{W}/\mbox{m } ^{\text{o}}$C&$\mbox{J}/\mbox{kg }^{\text{o}}$C & $\mbox{s}^{-1}$& ${^{\text{o}}\mbox{C}}^{-1}$&& MPa & &\\ \hline
          $80$ & $7860$ & $49.2$ & $473$ & $10^{-4}$ & $6.43\cdot 10^{-4}$ & $0.025$ & $600$ & $0.012$ & $0.107$\\ \hline
    \end{tabular}
  \end{center}
\end{table}

We have performed computations only for elastic, perfectly plastic problem. The computational domain is divided with 6941 nodes ($\Delta y= 10^{-6}$ m). Boundary conditions were used in the form \eqref{Eq8}, where $v_B=\dot{\varepsilon}_0H$. Initially velocity is given by the linear function $v(y,0)=\dot{\varepsilon}_0y$, the slab is free of stress $\tau_0(y)=0$ and the initial temperature distribution is taken as
\begin{equation} \label{eq:1.22}
T_0(y)=16.2{\left[ 1-4{\left( \frac{y}{H}-0.5\right)}^{2}\right]}^{9}\exp\left[ -20{\left( \frac{y}{H}-0.5\right)}^{2}\right].
\end{equation}
Such nonuniform distribution of initial temperature is used to trigger shear band formation at the center of the slab.
\begin{figure}[h]
\center
\includegraphics[width = 13 cm]{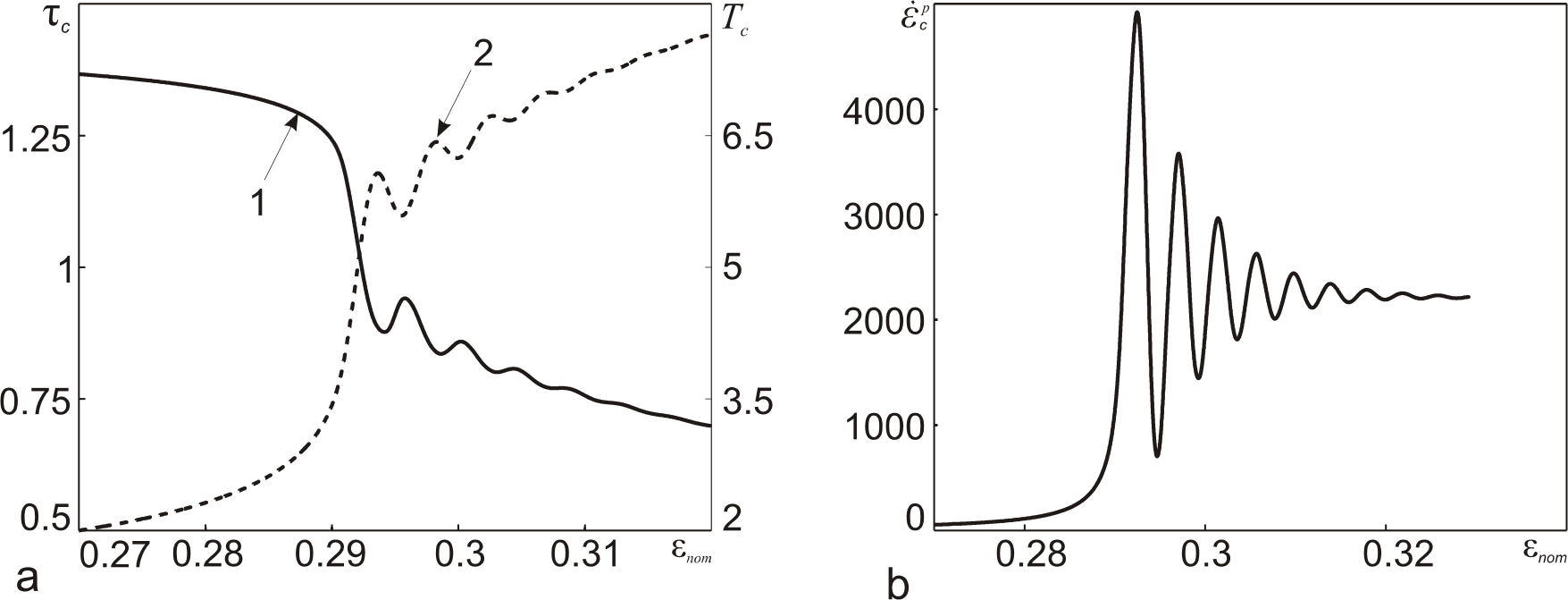}
\caption{(a) -- central stress and temperature (1,2), (b) -- central plastic strain rate ($\dot{\varepsilon}_0=750$ $s^{-1}$).}
\label{fig1}
\end{figure}

The evolution curves for the central values of stress, temperature and strain rate are shown on Fig. \ref{fig1}. We give only a narrow region where $\varepsilon_{nom} \in [0.27, 0.32]$. To compare our results with Walters results we scale the stress, temperature and strain rate by the values 600 MPa, 162 $^{\text{o}}$C and 750 s$^{-1}$, which are the same as in the work \cite{Walter1992}. From Fig. \ref{fig1} we see that the localization time for the current problem appears with the nominal plastic strain about 0.28, the maximum amplitude and the behavior of stress, temperature and plastic strain rate curves are all the same as in the Walters problem. Thus we may make a conclusion that our numerical methodology work correctly if we do not take into account strain hardening effects.

\subsection{Benchmark problem 2: Adiabatic shear band formation in OFHC copper taking into account strain hardening effects}

Let us present another benchmark problem, which was studied by Walter in \cite{Walter1992}. This problem differs from the previous one since it takes into account the process of strain hardening in material.

Let us consider the slab which made of OFHC copper. The slab is $3.18$ mm long and is deformed at strain rate of $\dot{\varepsilon}_0=330 \, \mbox{s}^{-1}$. The Litonski's flow law  is used to simulate the adiabatic shear bands formation
\begin{equation} \label{eq:1.53}
\tau=k(\psi)g(T)\left(1+m\cdot \ln(b|\dot{\varepsilon}^p|)\right),
\end{equation}
The thermal softening factor $g(T)$ and strain hardening factor $\kappa(\psi)$ are used in the form
\begin{equation} \label{eq:1.54}
g(T)={(1-aT)}^{3},
\end{equation}
\begin{equation}\label{eq:1.54_2}
\kappa(\psi)=\kappa_0\left(1+\left(\frac{\psi}{\psi_0}\right)^{n}\right).
\end{equation}
According to state Eq. \eqref{eq:1.53} we have found the formula for $\dot{\varepsilon}^p$ in the form
\begin{equation}
\dot{\varepsilon}^p=\frac{1}{b}\exp\left\{\frac{1}{m}\left(\frac{\tau}{\kappa(\psi)g(T)}-1\right)\right\}.
\end{equation}
The material parameters corresponding to OFHC copper are shown in Table \ref{tabl2}.
\begin{table}[h]
  \caption{Elastic, thermo-visco-plastic parameters of OFHC copper \cite{Walter1992}.}\label{tabl2}
  \begin{center}
    \begin{tabular}{c|c|c|c|c|c|c|c|c|c}
        \hline
$\mu\mbox{,}$& $\rho\mbox{,}$& $k\mbox{,}$ & $C\mbox{,}$ &$\dot{\varepsilon}_y$ &$a\mbox{,}$&$m$&$\kappa_0$&$\psi_0$&$n$\\
$\mbox{GPa}$&$\mbox{kg}/\mbox{m}^3$&$\mbox{W}/\mbox{m }^{\text{o}}$C&$\mbox{J}/\mbox{kg }^{\text{o}}$C&s$^{-1}$& ${^{\text{o}}\mbox{C}}^{-1}$& &MPa& & \\  \hline
          $45$&$8960$&$386$&$383$&$1$&$9.47\cdot 10^{-4}$ &$0.027$&$69$&0.261&0.32\\ \hline
    \end{tabular}
  \end{center}
\end{table}

The boundary conditions were the same as in the previous benchmark problem. The initial conditions for stress, velocity, temperature and strain hardening were taken in the following form
\begin{equation} \label{eq:1.56}
\begin{gathered}
\tau(y,t=0)=1+m\cdot \lg(b\dot{\varepsilon}_0),\\
v(y,t=0)=\dot{\varepsilon}_0y,\\
T(y,t=0)=0,\\
\psi(y,t=0)=0.
\end{gathered}
\end{equation}

\begin{figure}[h]
\center
\includegraphics[width = 13 cm]{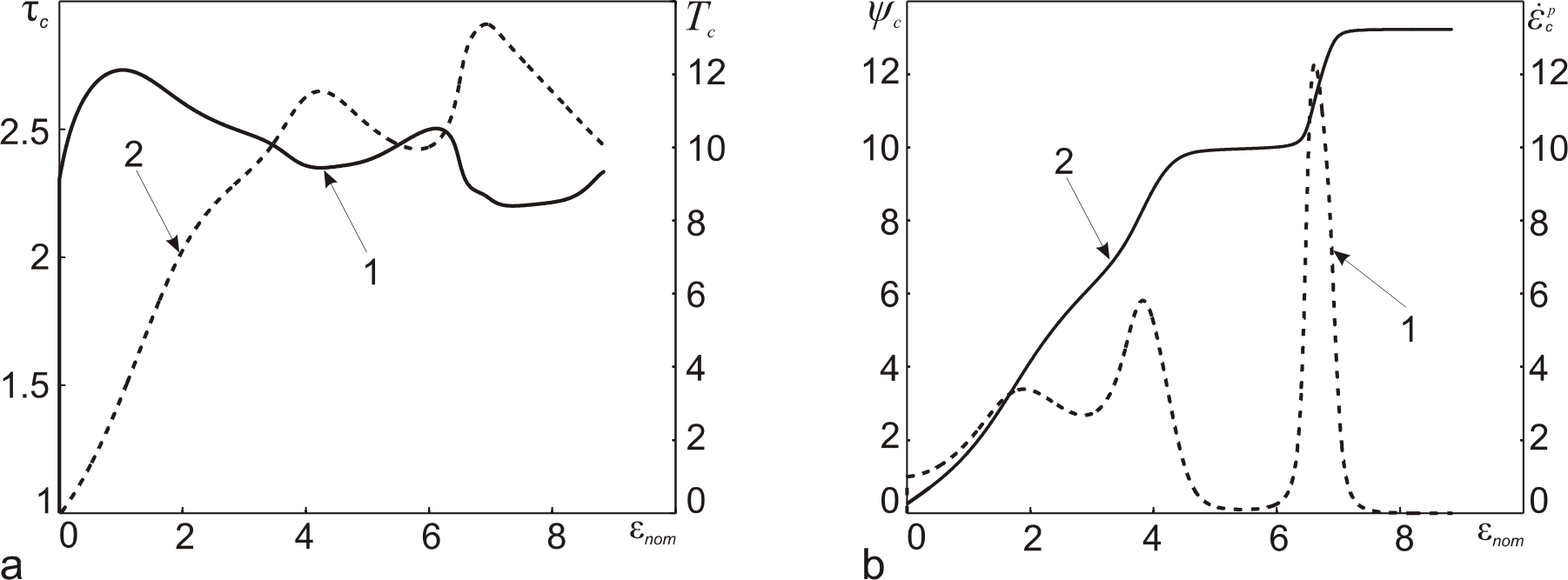}
\caption{(a) -- central stress and temperature (1,2), (b) -- central strain rate and strain hardening parameter (1,2) where ($\dot{\varepsilon}_0=330$ $s^{-1}$).}
\label{fig3}
\end{figure}
The slab was divided by the 3181 computational nodes such that $\Delta y=1 \, \mu$m. The stress $\tau$, temperature $T$ and plastic strain rate $\dot{\varepsilon}^p$ were scaled by the values $69$ MPa, $20$ ${^{\text{o}}}$C, $330$ s$^{-1}$. The calculation results are shown in Fig.\ref{fig3}. When we compare the results plotted in Fig. \ref{fig3} and Fig. 23 from \cite{Walter1992} we see that the evolution curves of stress, temperature, strain hardening and plastic strain rate are the same. Thus our numerical methodology works correctly even if we take into account the strain hardening effects.

\section{Numerical experiments on the formation of multiple adiabatic shear bands in OFHC copper and HY-100 steel}

Let us study the effect of strain-hardening on the formation of multiple adiabatic shear bands in materials. We consider two specimens which are made of OFHC copper and HY-100 steel. The height of the specimens was equal to 23.18 mm. We use the Litonski's flow law in the form
\begin{equation} \label{eq:1.58}
\tau=\kappa(\psi)g(T)\left(1+ \frac{\dot{\varepsilon}^p}{\dot{\varepsilon}_y}\right)^m,
\end{equation}
with thermal softening factor $g(T)$ described in Eq. (\ref{eq:1.54}) for OFHC copper and by \eqref{eq:1.20_1} for HY-100 steel. Thermo--physical parameters of steel and copper are given in Table \ref{tabl1}, \ref{tabl2} respectively. The function $\kappa(\psi)$ is taken as \eqref{eq:1.54_2} for both materials.

We assume that the initial temperature distribution over the samples is uniform where $T_0=0$ ${^{\text{o}}}$C. The velocity distribution over the height of the sample is linearly i.e. $v_0=\dot{\varepsilon}_0y$. Thus the initial conditions were used in the form
\begin{equation} \label{eq:1.59}
\begin{gathered}
\tau(y,0)=\tau_0(1+\xi), \quad \psi(y,0)=0,\\
T(y,0)=0,\quad v(y,0)=\dot{\varepsilon}_0y,\\
\dot{\varepsilon}^p(y,0)=\dot{\varepsilon}_0,\quad {\varepsilon}^p(y,0)=0,
\end{gathered}
\end{equation}
where $\xi$ is uniformly distributed random variable in $[0, 1]$. Because of inhomogeneities in the microstructure of the material the stress distribution over the sample becomes nonuniform. So the initial condition in this form properly describes the real physical problem. It should be noted that the similar way to generate ASB formation process was proposed by Batra et.al. earlier in \cite{Batra1988}. However authors used the periodic function with a known period to initiate growth of ASB. Thus the places, where ASB are formed, are known.

\begin{figure}[!htb]
\center
\includegraphics[width=13 cm]{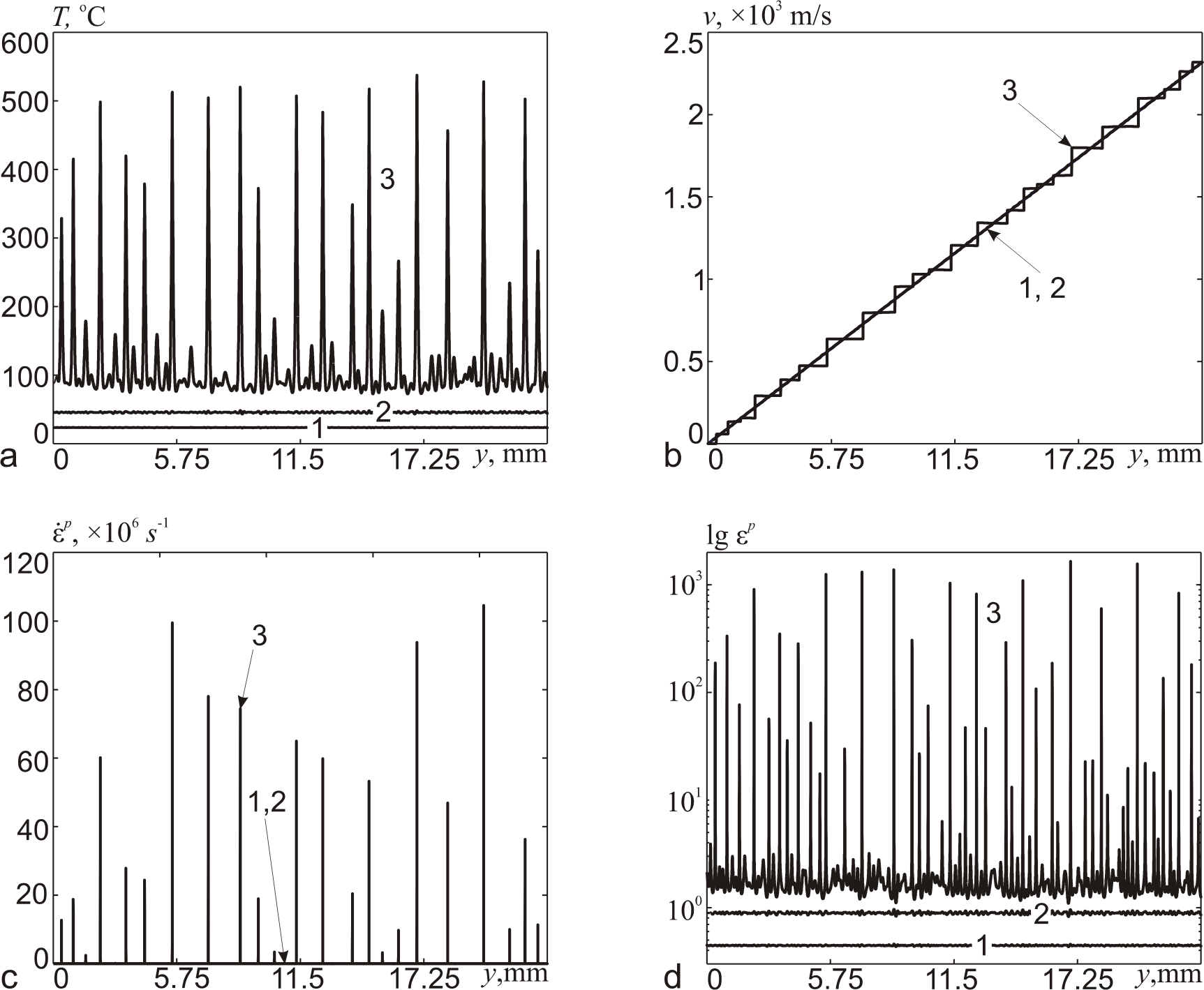}
\caption{Self-organization of adiabatic shear bands without strain hardening effect. Profiles of: (a) -- temperature $T$; (b) -- velocity $v$; (c) -- plastic strain rate $\dot{\varepsilon}^p$; (d) -- logarithm of plastic strain $\lg\varepsilon^p$; 1-3 -- $\varepsilon_{nom}=0.44, 1.33, 2.67$  ($\dot{\varepsilon}_0=10^5$ s$^{-1}$).}
\label{fig4}
\end{figure}

Fig. \ref{fig4} shows the distribution of temperature $T$, speed $v$, plastic strain $\varepsilon^p$ and plastic strain rate $\dot{\varepsilon^p}$ in the slab of OFHC copper at $\dot{\varepsilon}_0=10^5$ s$^{-1}$. At the first stage of the adiabatic shear bands formation process the temperature distribution over the height of the sample is uniform (Fig.\ref{fig4}a). The corresponding value of nominal strain is $\varepsilon_{nom}<1.33$.  Then there is a sharp increase in temperature of the sample in the areas of localization of adiabatic shear bands from $T\sim70$ $^{\text{o}}$C to $T\sim 500$ $^{\text{o}}$C. At the last stage the distribution of temperature becomes stationary on a whole computational domain. The velocity distribution over the height of the sample takes the form of stairs (Fig.\ref{fig4}b). Sharp jumps of the velocity are observed in the localization areas of shear bands. On Fig.\ref{fig4}c,d we plot the dependence of strain and plastic strain rate on the coordinate $y$ at different times. Fig.\ref{fig4}c shows that when $\varepsilon_{nom}<1.33$ there is an uniform distribution of strain $\varepsilon$ in the specimen and $\dot{\varepsilon}^p\approx\dot{\varepsilon}_0$. In the case when $\varepsilon_{nom}>1.33$ sharp jumps of strain (from $\varepsilon^p=0.4$ to $\varepsilon^p=100$ and more) appear in the localization areas. From physical point of view the value of plastic strain $\varepsilon^p$ cannot be 100 or larger since the material point will fail even at a lower plastic strain. In the case of HY-100 steel the distribution of strain and temperature is uniform when $\varepsilon_{nom}< 1.25$. If $\varepsilon_{nom} \geqslant 1.25$ we observe a sharp jump of the temperature in localization areas from $T\sim200$ $^{\text{o}}$C to $T\sim 1000$ $^{\text{o}}$C. Numerical experiments show that the localization area of each shear band in OFHC copper and HY-100 steel is very small. The mechanical band width for OFHC copper approximately equals $10\, \mu$m and $1\, \mu$m for HY-100 steel. The initial value of strain rate does not effect on the band width as well as the refinement of the mesh.

\begin{figure}[!htb]
\center
\includegraphics[width=13 cm]{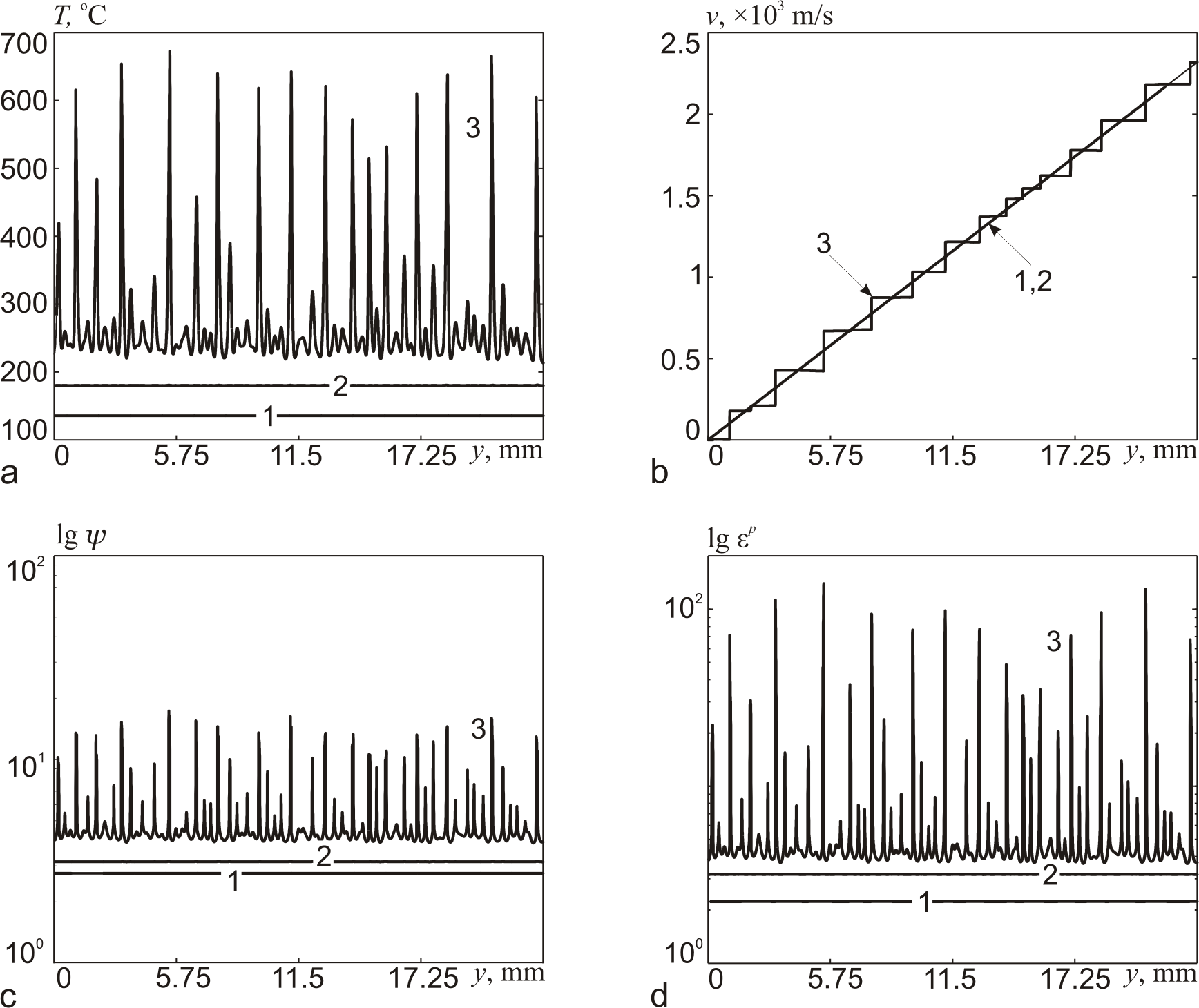}
\caption{Self-organization of adiabatic shear bands with strain hardening effect. Profiles of: (a) -- temperature, $T$; (b) -- velocity of material particles, $v$; (c) -- logarithm of strain hardening variable $\psi$, $\dot{\varepsilon}^p$; (d) -- logarithm of plastic strain $\lg \varepsilon^p$; 1-3 -- $\varepsilon_{nom}=2.23, 3.12, 6.24$  ($\dot{\varepsilon}_0=10^5$ s$^{-1}$).}
\label{fig5}
\end{figure}

Fig. \ref{fig5} shows the profiles of temperature, velocity, logarithm of strain hardening variable and plastic strain taking into account the strain hardening process. This process leads to an increase of the material strength and so it leads to an increase of localization time. For OFHC copper the value of the temperature in ASB regions increase on $\sim 100^{\text{o}}$C and for HY-100 steel the temperature in ASB regions increase on $\sim 250^{\text{o}}$C. We have noted earlier, that $\psi$ and $\varepsilon^p$ has the same physical interpretation, but their values in localization areas of ASB (which coincides) differ approximately in five times for OFHC copper and in three times for HY-100 steel (Fig. \ref{fig5}).

\begin{figure}[!htb]
\center
\includegraphics[width=13 cm]{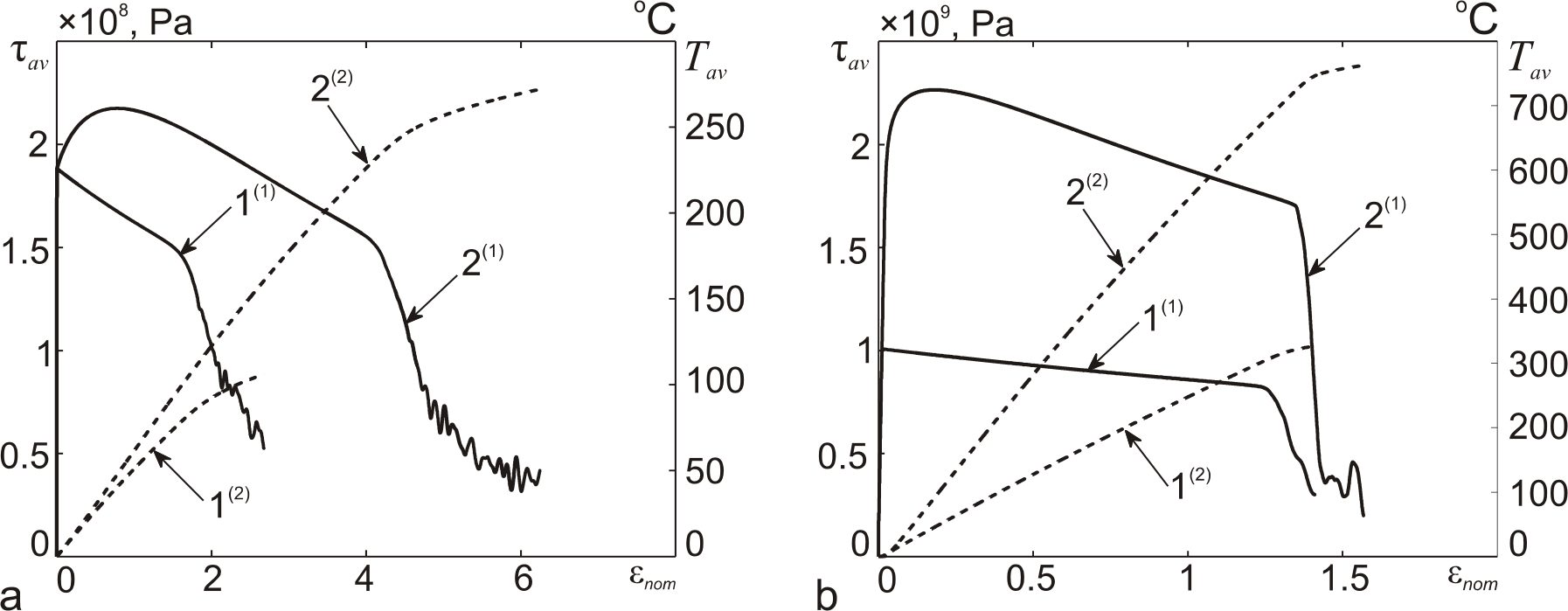}
\caption{Distribution of average stress(solid line) and temperature(dashed line) for OFHC copper (a) and HY-100 steel (b). $1^{(1, 2)}$  -- average stress and temperature without strain hardening, $2^{(1, 2)}$ -- average stress and temperature with strain hardening. ($\dot{\varepsilon}_0=10^5$ s$^{-1}$).}
\label{fig6}
\end{figure}

\begin{figure}[h]
\center{\includegraphics[width=13 cm]{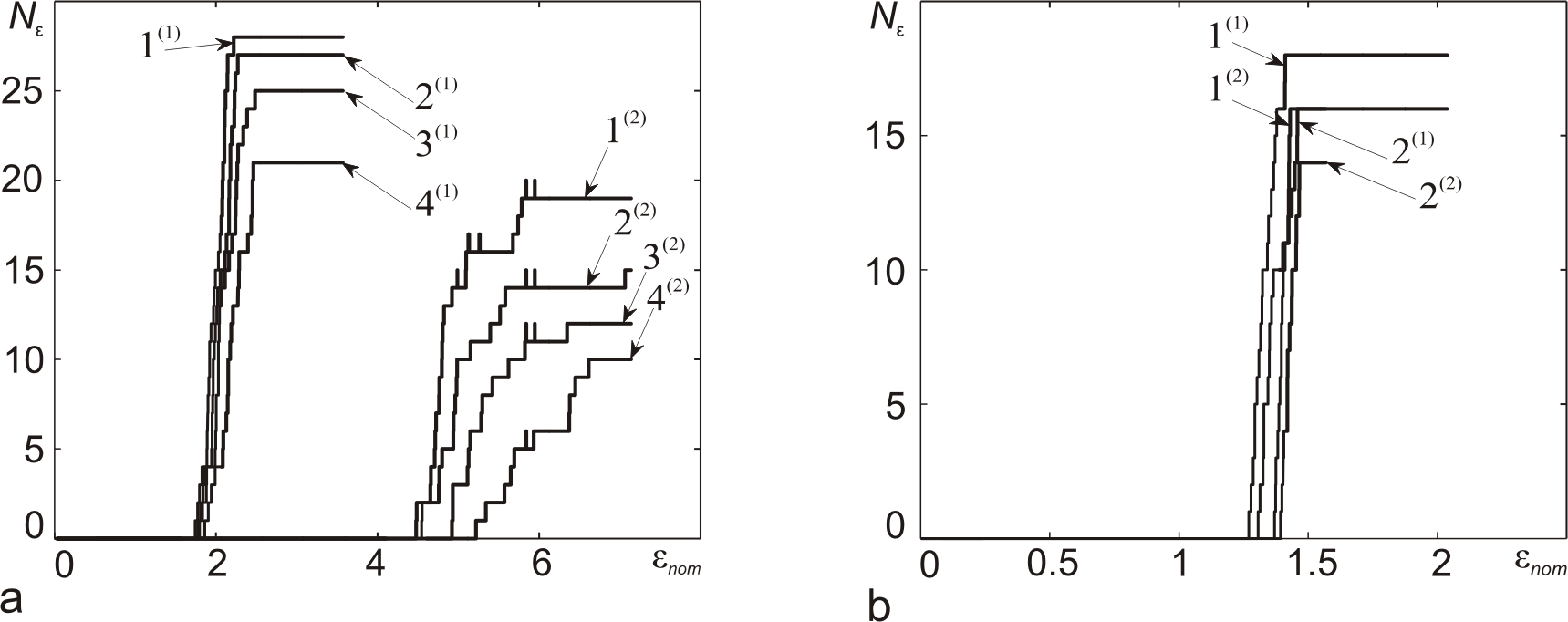}}
\caption{Number of formed adiabatic shear bands $N_{\varepsilon}$ counted for OFHC copper (a) and HY-100 steel (b). For (a): $1^{(1,2)}-4^{(1,2)}$ -- without and with strain hardening effect ($\varepsilon_{tr}=30, 40, 50, 80$). For (b): $1^{(1,2)}-2^{(1,2)}$ -- without and with strain hardening effect ($\varepsilon_{tr}=30, 80$) ($\dot{\varepsilon}_0=10^5$ s$^{-1}$).}
\label{fig7}
\end{figure}

Evolution of the average values of stress $\tau_{av}$ and temperature $T_{av}$ at $\dot{\varepsilon}_0=10^5$ $s^{-1}$ are presented in Fig.\ref{fig6}, taking into account the strain hardening effect and not. The average values $\tau_{av}$, $T_{av}$ are calculated using the following formulas \cite{Zhou2006b}
\begin{equation}
\begin{gathered}
T_{av}=\frac{1}{J+1}\sum_{j=0}^{J}T_j, \quad \tau_{av}=\frac{1}{J+1}\sum_{j=0}^{J} \tau_j.
\end{gathered}
\end{equation}
Based on the obtained dependencies we can define the localization time like the time when $d\tau/d\varepsilon_{nom}$ tends to its maximum. Usually at this moment $\tau_{av}$ lies between $80 - 90\%$ from its maximum value. It should be noted that the following criterion implies that shear bands formed simultaneously. So their collective behavior can be considered as the behavior of one band with thermo-physical characteristics presented on Fig. \ref{fig6}. This assumption can be done since most shear bands form approximately at the same time, see Fig. \ref{fig7}. Numerical experiments showed that value of localization time decreases with increasing of the initial strain rate and varies from $70 \, \mu$s to $13.5 \, \mu$s for HY-100 steel and from $204 \, \mu$s to $43 \, \mu$s for OFHC copper at $2\times 10^4\leqslant \dot{\varepsilon}_0\leqslant10^5$ s$^{-1}$. Let us note that nominal strain for HY-100 approximately equal to $\varepsilon_{nom}=1.39$  and $\varepsilon_{nom}=4.2$ for OFHC copper. We note, that it is possible that if the strength of perturbations will be widely different this estimate cannot be used. However, this assumption gives good results even if we increase the amplitude of initial disturbances in twice. From Fig. \ref{fig6} we see that the effect of strain hardening significantly influences on the evolution of the stress-temperature curves and on the value of the localization time. The average temperature at localization time in OFHC copper is approximately about $T_{av}=80 \, ^{\text{o}}$C and if we take into account strain hardening the average temperature is  $T_{av}= 182\, ^{\text{o}}$C. As for HY-100 steel, the average temperature is about $T_{av}=256 \, ^{\text{o}}$C without strain hardening effect and if we take the hardening process into account average temperature becomes $T_{av}=725 \, ^{\text{o}}$C. It should be noted that for OFHC copper there is no significant difference between the average value of stress with and without strain hardening effect at the moment of localization however for steel the values of stress differ twice (Fig.\ref{fig6}).

\begin{figure}[h]
\center{\includegraphics[width=13 cm]{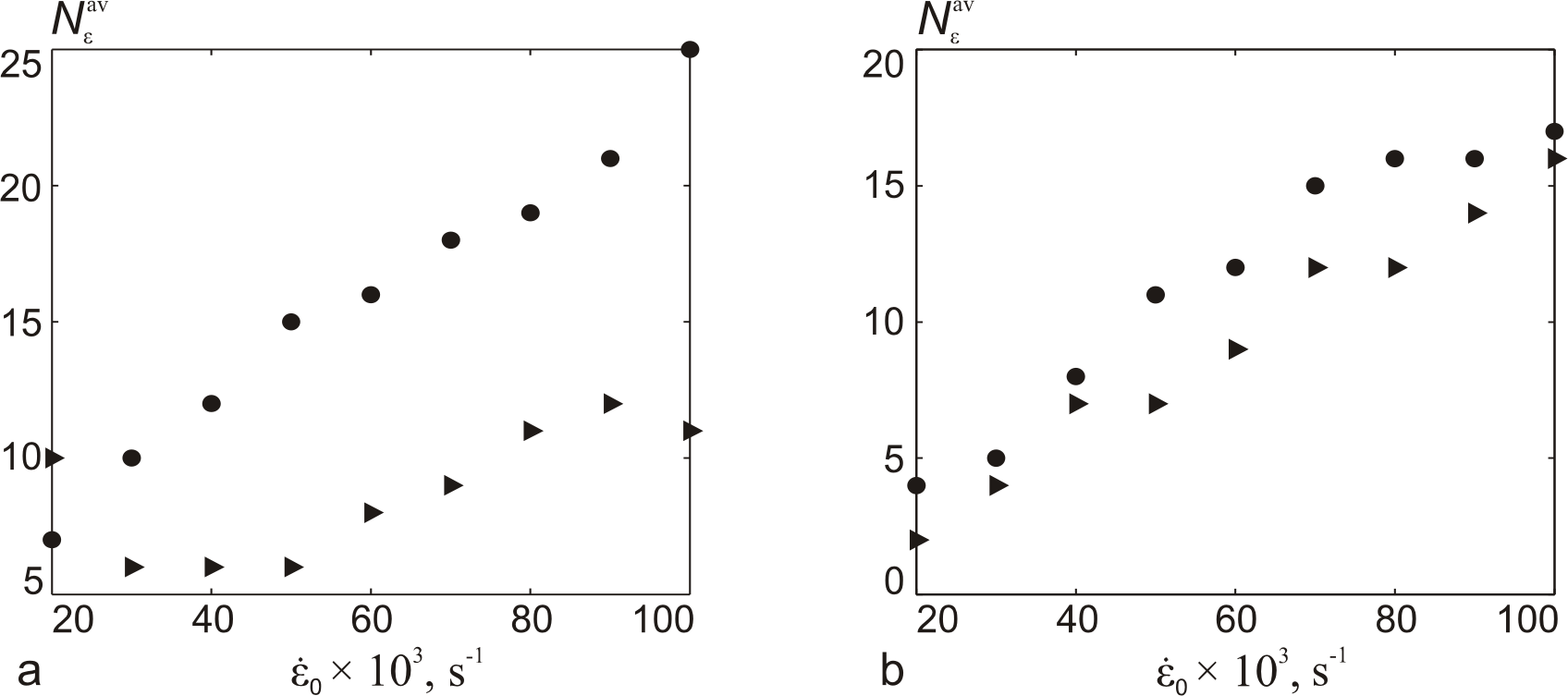}}
\caption{Dependence of average value of $N_{\varepsilon}^{av}$ from $\dot{\varepsilon}_{0}$ for OFHC copper (a) and HY-100 steel (b). Dots -- without strain hardening effect, triangles -- with strain hardening effect. (a) -- $\varepsilon_{tr}=50$, (b) -- $\varepsilon_{tr}=80$. ($\dot{\varepsilon}_0=10^5$ s$^{-1}$).}
\label{fig8}
\end{figure}

Adiabatic shear bands are the regions in which high temperatures and deformations emerge for a very short period of time. That is why we can calculate the number of bands using two different criteria. According to the first one the number of bands $N_{\varepsilon}$ equal to the number of local maxima of strain $\varepsilon^p$, which exceed a threshold strain $\varepsilon_{tr}$ \cite{Zhou2006b, KKR2011}. On the other hand we can calculate the number of bands $N_{T}$ as a number of local maxima of temperature which exceed a threshold value $T_{tr}$. Note that in strain criteria one can use the value of $\psi$ but the location of the maxima for $\psi$ and $\varepsilon^p$ are the same thus there is no matter what we use. However, it is more convenient to use $\varepsilon^p$, because we can compare the number of bands formed in the case when we take $\psi$ into account and not using the same threshold strains $\varepsilon_{tr}$. Numerical experiments show that strain criteria more useful then temperature one \cite{KKR2011}. Let us explain this fact. The evolution of the value of $N_{\varepsilon}$ with time can be divided on three stages. At the first stage the value of $N_{\varepsilon}=0$. The second stage corresponds to the beginning of the localization process. The value of  $N_{\varepsilon}$ increases sharply. Finally, on the third stage the value of $N_{\varepsilon}$ becomes constant. The example of such evolution is plotted in Fig. \ref{fig7} or one can find it in works \cite{Zhou2006b, KKR2011}. However the behavior of $N_{T}$ does not obey to this scenario. When the localization process takes place the value of $N_{T}$ tends to its maximum ($\max N_{T} > \max N_{\varepsilon}$) and then starts to decrease stepwise \cite{KKR2011}. After a while the value of $N_{T}$ becomes approximately equal to $N_{\varepsilon}$. Thus we use the deformation criteria to define the number of bands in this work. Moreover for each value of initial strain rate $\dot{\varepsilon}_0$ , we have performed numerical simulation with several distributions of random disturbances of defects and then calculate the average value of $N_{\varepsilon}^{av}$ for each value of $\dot{\varepsilon}_0$ .  The dependence of $N_{\varepsilon}^{av}$ from the initial plastic strain rate $\dot{\varepsilon}_0$ at threshold strain $\varepsilon_{tr}=50$ for OFHC copper and $\varepsilon_{tr}=80$ for HY-100 steel for cases with and without strain hardening is illustrated in Fig.\ref{fig8}. From Fig.\ref{fig8} we see that the number of adiabatic shear bands increases with increasing the value of initial strain rate.

\begin{figure}[h]
\center{\includegraphics[width=13 cm]{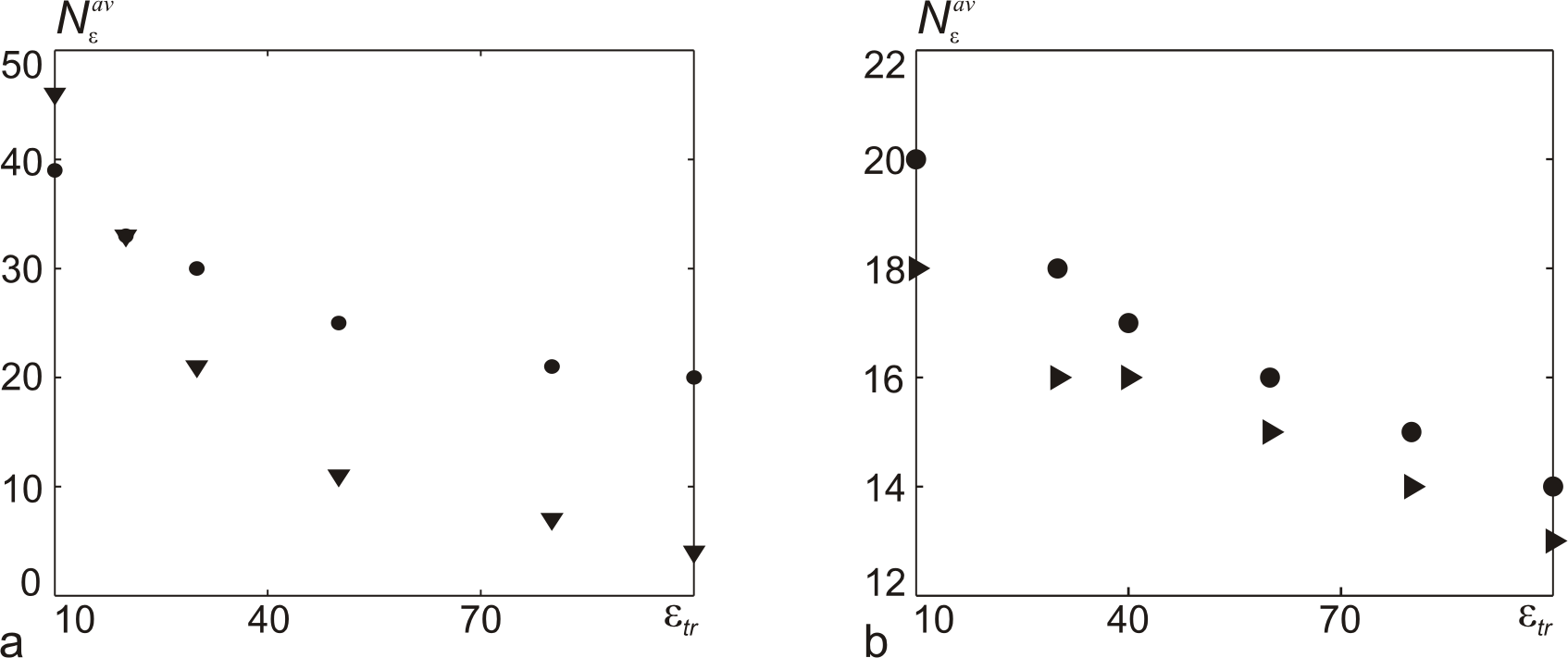}}
\caption{Dependence of $N_{\varepsilon}$ from threshold strain $\varepsilon_{tr}$ for OFHC copper (a) and HY-100 steel (b). Dots -- without strain hardening effect, triangles -- with strain hardening effect. ($\dot{\varepsilon}_0=10^5$ $s^{-1}$).}
\label{fig9}
\end{figure}

From Fig. \ref{fig7} and Fig. \ref{fig8} follows that if we take into account the processes of strain hardening the total number of shear bands decreases. For OFHC copper the amount of shear bands decreases approximately on $30-40 \%$, for HY-100 steel on $12-25 \%$. Note, that above mentioned relations do not depend on threshold strain. However the value of $N_{\varepsilon}^{av}$ depends on the value of $\varepsilon_{tr}$. The following dependencies for copper and steel are illustrated in Fig.\ref{fig9} for the case when $\dot{\varepsilon}_0=10^5$ s$^{-1}$.

It is known that the adiabatic shear bands formation leads to the appearance of cracks in their localization areas. Therefore the estimation of the distances between ASB is very important problem. Using the theoretical and numerical approach in works \cite{Wright-Ockendon, Grady-Kipp, Zhou2006b} authors proposed the convenient formulas for the estimation of the distance between ASB. These formulas take the form
\begin{equation} \label{eq:1.60}
\begin{gathered}
L_{GK}=2{\left(\frac{9 \kappa C}{k_0a^2{\dot{\varepsilon}_0}^3} \right)}^{1/4}, \quad L_{WO}=2 \pi {\left(\frac{m^3 \kappa C}{k_0a^2{\dot{\varepsilon}_0}^3 (\dot{\varepsilon}_0/\dot{\varepsilon}_y)^m} \right)}^{1/4},\\
L_{ZWR}=\sqrt{\frac{2}{2+m}}\pi m^{3/2} {\left(\frac{c}{a{\dot{\varepsilon}_0}^2}\right)}^{1/2}{\left[1+{\left(1+8\frac{k\dot{\varepsilon}_0}{m^3ck_0}\right)}^{1/2} \right]}^{1/2}.
\end{gathered}
\end{equation}

\begin{figure}[!htb]
\center{\includegraphics[width=13 cm]{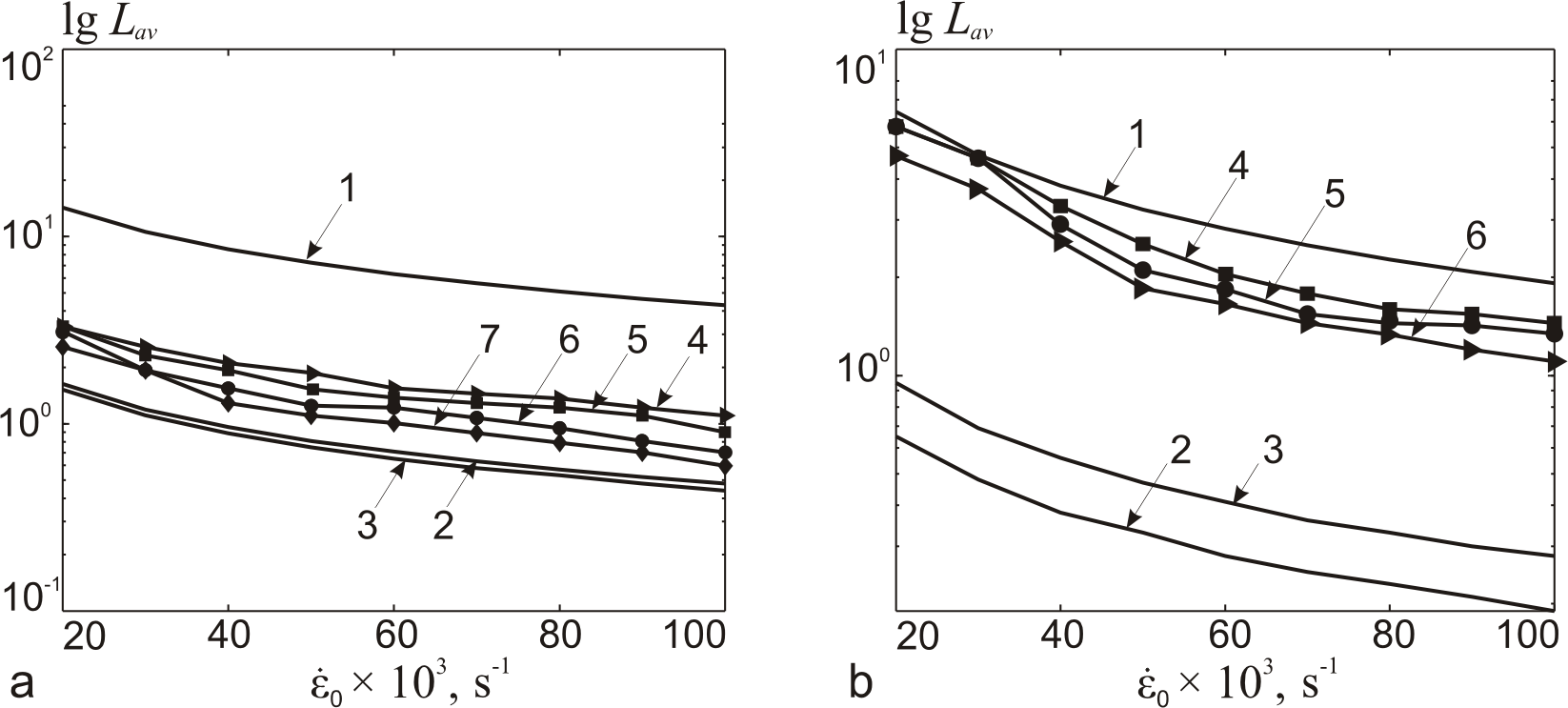}}
\caption{Dependence of logarithm of average distance between ASB $\lg L_{av}$ from initial strain rate $\dot{\varepsilon}_0$ for OFHC copper (a) and HY-100 steel (b) taking into account strain hardening. Curves 1-3 corresponds to $\lg L_{GK}, \lg L_{WO}$ and $\lg L_{ZWR}$ estimates. For (a): $\varepsilon_{tr}=80, 50, 20, 10$ (curves 4-7). For (b): $\varepsilon_{tr}=80, 50, 10$ (curves 4-6).}\label{fig10}
\end{figure}

In the present work to calculate the distance between the ASB we use the formula
\begin{equation} \label{eq:1.61}
L_{av}=\frac{H}{m}\sum_{i=1}^{m}\frac{1}{N_{\varepsilon, i}},
\end{equation}
where $m$ is a number of calculations corresponding to specific value of initial strain rate $\dot{\varepsilon}_0$ with different distribution of initial disturbances. It should be noted that using the formula \eqref{eq:1.61} we obtain the average distance between ASB. This estimation seems acceptable since the distance between localization areas can be considered the same (Fig. \ref{fig4}, \ref{fig5}) for both materials and the process of ASB formation is quasiperiodic. Moreover this fact is confirmed by works of Nesterenko and et.al. \cite{Nesterenko1998, Nesterenko2002, Nesterenko2004}. Also we have calculated the standard deviation and relative error in shear band spacing. From Fig. \ref{fig10} we see that the value of average band spacing decreases with increasing of initial strain rate. The same behavior is observed for absolute value of standard deviation $s$. For example in the case of HY-100 steel at $\dot{\varepsilon}_0=10^5$ s$^{-1}$ average band spacing equals $L_{av}=1.34$ with $s=0.104$ and if $\dot{\varepsilon}_0=70 \times 10^3$ s$^{-1}$ then $L_{av}=1.97$ with $s=0.29$. In the case of OFHC copper band spacing equals $L_{av}=0.957$ with $s=0.036$ and $L_{av}=1.289$ with $s=0.047$ for $\dot{\varepsilon}_0=10^5 $ and $70 \times 10^3$ s$^{-1}$ respectively. Here we take  $\varepsilon_{tr}=50$. In turn, the value of relative error varies on the interval $3 - 6 \%$  for OFHC copper and on $8-18 \%$ for HY-100 steel. Numerical experiments show that these values do not depend on initial strain rate. Let us note that the main advantage of the estimation \eqref{eq:1.61} is the fact that we can use it when specific heat $k$, thermal conductivity $C$ and shear module $\mu$ are functions of temperature.

Fig.\ref{fig10} shows the comparison of numerical and theoretical estimates of the distance between ASB depending on the initial strain rate $\dot{\varepsilon}_0$ at different threshold strains $\varepsilon_{tr}$ taking into account the strain hardening process. The results of the numerical simulation lie between the estimates of $L_{WO}, L_{ZWR}$ and the estimate of $L_{GK}$, i.e. $L_{GK}\leq L_{av}\leq L_{WO}, L_{ZWR}$. It turned out that for HY-100 steel the results are closer to $L_{GK}$, however for OFHC-copper the results are closer to $L_{WO}, L_{ZWR}$. Note that logarithm of average distance between ASB decreases linear with increasing of initial strain rate $\dot{\varepsilon}_{0}$.

\section{Conclusions}
The process of multiple adiabatic shear bands formation in OFHC copper and HY-100 steel is considered. The mathematical model described the process of ASB formation is formulated taking into account the strain hardening factor. To solve this problem we present the finite difference scheme that is based on Courant -- Isaacson -- Rees scheme. This algorithm has several advantages. First of all, it allows observing numerically the process of ASB formation from the initial to the final stage of localization. Its efficiency and accuracy were analyzed on two benchmark problems by Walter \cite{Walter1992}. Both tests show that our results are identical to those obtained in \cite{Walter1992}. Secondly, this approach can be used when the shear modulus $\mu$, specific heat $C$ and thermal conductivity $k$ are functions of temperature. For example the algorithm proposed in \cite{Zhou2006a, Zhou2006b} cannot be used in this case. Thirdly, it allows taking into account strain hardening effect. Since our approach based on the explicit finite-difference scheme it is easier in realization than numerical approaches based on finite element method \cite{Walter1992, BatraKim1990}. Moreover, in order to accelerate computation the algorithm can be easily parallelized.

Using the above mentioned algorithm we have provided the numerical simulation of adiabatic shear bands formation taking into consideration the strain hardening effect. It was shown that this effect changes the quantitative characteristics of ASB formation process. The temperature in ASB regions increase approximately on $\sim 100^{\text{o}}$C for OFHC copper and on $\sim 250^{\text{o}}$C for HY-100 wherein the average temperature of the specimen increase from $80^{\text{o}}$C to $182^{\text{o}}$C and from $256^{\text{o}}$C to $725^{\text{o}}$C for copper and steel respectively. Numerical experiments show that the strain hardening also influence on the total number of shear bands formed. The number of ASB decrease approximately on $30-40 \%$ for OFHC copper  and on $12-25 \%$ for HY-100 steel. It was shown that the average distance between ASB in HY-100 steel is closer to Grady--Kipp estimate but for OFHC copper the distance tends to Wright--Oscendon and Wright--Zhou--Ramesh estimates.

\section{Acknowledgement}
This work was supported by Russian Science Foundation, project to support research carried out by individual research groups No. 14-11-00258


\begin{thebibliography} {99}

\addcontentsline{toc}{section}{References}

\bibitem{Schneider_2004} Schneider, J., Nunes, J.A., 2004. Characterization of plastic flow and resulting microtextures in a friction stir weld. Metall. Mater. Trans. B. 35, 777--783.

\bibitem{Seidel2001} Seidel, T., Reynolds, A., 2001. Visualization of the material flow in aa2195 friction stir welds
using a marker insert technique. Metall. Mater. Trans. 32A, 2879--2884.

\bibitem{Moss1981} Moss, G., 1981. Shear strains, strain rates, temperature changes in adiabatic shear bands. Shock Waves and High Strain Rate Phenomena in Metals / Ed. by L. Meyers, L. Murr. 299--312.

\bibitem{Rittel2005} Rittel, D. 2005 Adiabatic shear failure of a syntactic polymeric foam. Materials Letters. 59, 1845--1848.

\bibitem{Shockey2007} Shockey, D.A., Simons, J.W., Brown, C.S., Kobayashi, T., 2007. Shear failure of inconel 718 under dynamic loads. Experimental Mechanics. 47, 723--732.

\bibitem{Nesterenko1995} Nesterenko, V.F., Meyers, M.A., Wright, T.W., 1995. Collective behaviour of shear bands. In.: Murr, L.E., Staudhammer, K.P., Meyers, M.A. (Eds.) Metallurgical and Materials Application of Shoch-wave and High-Strain-Rate Phenomena. Elsivier Science, Amsterdam, 397--404.

\bibitem{Nesterenko1998} Nesterenko, V.F., Meyers, M.A., Wright, T.W. 1998. Self-organization in the initiation of adiabatic shear
bands. Acta Materialia 46, 327--340.

\bibitem{Nesterenko2002} Xue, Q., Meyers, M.A., Nesterenko, V.F., 2002. Self-organization of shear bands in titanium and Ti--6Al--4V
alloy. Acta Materialia 50, 575--596.

\bibitem{Nesterenko2004} Xue, Q., Meyers, M.A., Nesterenko, V.F. 2004. Self organization of shear bands in stainless
steel. Materials Science and Engineering A 384, 35--46.

\bibitem{Grady-Kipp} Grady, D.E., Kipp, M.E. 1987. The growth of unstable thermoplastic shear with application to steady-wave shock compression in solids. J. Mech. Phys. Solids 35, 95--118.

\bibitem{Wright-Ockendon} Wright, T.W., Ockendon, H. 1996. A scaling law for the effect of inertia on the formation of adiabatic shear bands. Int. J. Plasticity 12, 927--934.

\bibitem{Molinari} Molinari, A., 1997. Collecive behaviour and spacing of adiabatic shear bands. J. Mech. Phys. Solids 45, 1551--1575.

\bibitem{BatraChen1999} Batra, R.C., Chen, L., 1999. Shear band spacing in gradient-dependent thermoviscoplastic materials. Computational Mechanics 23, 8--19.

\bibitem{BatraChen2000} Chen, L., Batra, R.C., 2000. Microstructural effects on shear instability and shear band spacing. Theoret. Appl. Fract. Mech. 34, 155--166.

\bibitem{BatraWei2006} Batra, R.C., Wei, Z.G., 2006. Shear band spacing in thermoviscoplastic materials. International Journal of Impact Engineering 32, 947--967.

\bibitem{BatraWei2007} Batra, R.C., Wei, Z.G., 2007. Instability strain and shear band spacing in simple tensile/compressive deformations of thermoviscoplastic materials. International Journal of Impact Engineering 34,448--463.

\bibitem{Zhou2006a} Zhou, F., Wright, T.W., Ramesh, K.T., 2006. A numerical methodology for investigating the formation of adiabatic shear bands. J. Mech. Phys. Solids 54, 904--926.

\bibitem{Zhou2006b} Zhou, F., Wright, T.W., Ramesh, K.T., 2006. The formation of multiple adiabatic shear bands. J. Mech. Phys. Solids 54, 1376--1400.

\bibitem{DiLellio1997} DiLellio, J.A., Olmstead, W.E., 1997. Temporal evolution of shear band thickness. J. Mech. Phys. Solids 45, 345--359.

\bibitem{DiLellio1998} DiLellio, J.A., Olmstead, W.E., 1998. Numerical solutions of shear localization in a finite slab. Mech. Mater. 29, 71--80.

\bibitem{DiLellio2003} DiLellio, J.A., Olmstead, W.E., 2003. Numerical solution of shear localization in Johnson--Cook materials. Mechanics of Materials 35, 571--580.

\bibitem{Bayliss1994} Bayliss, A., Belytschko, T., Kulkarni, M., Lott-Crumpler, D.A., 1994. On the dynamics and the
role of imperfections for localization in thermal-viscoplastic materials. Model. Simul. Mater. Sci. Engi. 2,
941--964.

\bibitem{BatraKim1990} Batra, R.C., Kim, C.H., 1990. The interection among adiabatic shear bands in simple and dipolar materials. Int. J. Engn Sci. 28, 926 -- 942.

\bibitem{Doridon2004} Daridon, L., Oussouaddi, O., Ahzi, S., 2004. Influence of the material constitutive models on the
adiabatic shear band spacing: MTS, power law and Johnson–Cook models. Int. J. Solids Struct. 41, 3109--3124.

\bibitem{Edwards1998}  Edwards, D.A., French, D.A., 1998. Asymptotic and computational analysis of large shear deformations of a thermoplastic material. Siam J. Appl. Math. 59, 700--724.

\bibitem{Walter1992} Walter, J.W., 1992. Numerical experiments on adiabatic shear band formation in one dimension. Int. J. Plasticity 8, 657--693.

\bibitem{Batra1988} Known, Y.W., Batra, R.C., 1988. Effect of multiple initial imperfections on the initiation and growth of adiabatic shear bands in nonpolar and dipolar materials. Int. J. Engn Sci. 26, 1177 -- 1187.

\bibitem{Wright1985} Wright, T.W., Batra, R.C., 1985. The Initiation and Growth of Adiabatic Shear Bands. Int. J. Plasticity 1, 205--212.

\bibitem{Wright1987} Wright, T.W., Walter, J{\tiny R}., J.W., 1987. On Stress Collapse in Adiabatic Shear Bands. J. Mech. Phys. Solids 35, 701--720.

\bibitem{JohnsonCook} Johnson, G.R., Cook, W.H., 1981. A constitutive model and data for metals subjected to larg strains, high strain rates, and high temperatures. Proceedings of the seventh international symposium on ballistics, the hague, the netherlands, 541--548.

\bibitem{KKR2011} Koshkin, V.I., Kudryashov, N.A., Ryabov, P.N. 2011. Modeling of the quasiperiodic processes of adiabatic shear bands formation. Mathematical modeling 23, 117--132. (in Russian)

\bibitem{Koshkin2010} Koshkin, V.I., Kudryashov, N.A., Ryabov, P.N. 2010. Numerical simulation of adiabatic shear bands formation under deformations. Nuclear physics and engeenering 1, 465--474. (in Russian)

\bibitem{Batra2006} Batra, R.C., Wei, Z.G. 2006. Shear bands due to heat flux prescribed at boundaries. Int. J. Plasticity 22, 1--15.

\bibitem{Chm_Kulikovskii} Kulikovskii, A.G., Pogorelov, N.V., Semenov, A.Yu., 2000. Mathematical aspects of numerical solution of hyperbolic systems. In: Brezis, H., Douglas, R.G., Jeffrey A. (Eds.), Chapman and Hall/CRC monographs and surveys in pure and applied mathematics, 560.
\end{thebibliography}
\end{document}